\documentclass[aps,prl,reprint,superscriptaddress,showpacs,preprintnumbers,nofootinbib]{revtex4-1}
\usepackage{graphicx}
\usepackage{bm}
\usepackage{epsfig}
\usepackage{amsmath}
\usepackage{hyperref}
\usepackage{color}
\usepackage{float}
\usepackage{url}

\newcommand{\afb}{A_{\rm FB}}

\begin{document}

\title{Precise QCD predictions on top quark pair production \\
mediated by massive color-octet vector boson at hadron colliders}
\author{Hua Xing Zhu}
\affiliation{Department of Physics and State Key Laboratory of Nuclear Physics
and Technology, Peking University, Beijing 100871, China}
\author{Chong Sheng Li}\email{csli@pku.edu.cn}
\affiliation{Department of Physics and State Key Laboratory of Nuclear Physics
and Technology, Peking University, Beijing 100871, China}
\affiliation{Center for High Energy Physics, Peking University, Beijing 100871, China}
\author{Ding Yu Shao}
\affiliation{Department of Physics and State Key Laboratory of Nuclear Physics
and Technology, Peking University, Beijing 100871, China}
\author{Jian Wang}
\affiliation{Department of Physics and State Key Laboratory of Nuclear Physics
and Technology, Peking University, Beijing 100871, China}
\author{C.-P. Yuan}\email{yuan@pa.msu.edu}
\affiliation{Center for High Energy Physics, Peking University, Beijing 100871, China}
\affiliation{Department of Physics and Astronomy, Michigan State University,
East Lansing, MI 48824, USA}


\begin{abstract}
We present a theoretical framework for systematically calculating next-to-leading order (NLO) QCD effects to various experimental observables in models with massive COVB in a model independent way at hadron colliders. Specifically, we show the numerical results for the NLO QCD corrections to total cross sections, invariant mass distribution and $\afb$ of top quark pairs production mediated by a massive COVB in both the fixed scale~(top quark mass) scheme and the dynamical scale~(top pair invariant mass) scheme. Our results show that the NLO QCD calculations in the dynamical scale scheme is more reasonable than the fixed scheme and the naive estimate of the NLO effects by simple rescaling of the LO results with the SM NLO K-factor is not appropriate.
\end{abstract}

\maketitle

In many extensions of the Standard Model~(SM), massive Color-Octet Vector Boson~(COVB) is necessarily engaged at the TeV scale, for example, in the top-color~\cite{Hill:1991at}, warped~(RS) or universal extra dimensions~\cite{Randall:1999ee,Appelquist:2000nn}, technicolor~\cite{Lane:1991qh} and chiral color models~\cite{Pati:1974zv}. In all these cases, the COVB could have large impacts on the interaction of top quarks, which are being copiously produced at the CERN Large Hadron Collider~(LHC). With a large sample of $t\bar{t}$ data, the CDF collaboration at the Tevatron has recently reported an observation of a large Forward-Backward asymmetry~($A_{\rm FB}$) in $t\bar{t}$ production, $\afb^{\rm tot}= 0.158\pm0.075$, compared with the SM prediction $0.058\pm 0.009$~\cite{Kuhn:1998kw,Antunano:2007da,mcfm}. The disagreement is more profound in the region of large $t\bar{t}$ invariant mass, where CDF reported $\afb ({m_{t\bar{t}}>450{\rm GeV}})=0.475\pm 0.114$, and the SM gives $0.088\pm 0.013$~\cite{Aaltonen:2011kc}. This leads to a more than $3.4\,\sigma$ deviation from the SM prediction~\cite{Aaltonen:2011kc}. Similarily, the DO\!\!\!\!\slash\, collaboration~\cite{Abazov:2011rq} has also reported the total $A_{\rm FB}$ to be $\afb^{\rm tot}=0.196\pm 0.065$ using $5.4{\rm fb}^{-1}$ of data. Furthermore, DO\!\!\!\!\slash\ has also measured the charge asymmetry of the leptons from top decay, $A_{\rm FB}^l = 0.152\pm 0.04$. These results have generated extensive theoretical studies on this observable in various models beyond the SM. Among these,
models with massive COVB are in particular attractive, {\it c.f.}, Ref.~\cite{Kamenik:2011wt} and references therein.
There have also been substantial efforts in searching for the signal of COVB at the Tevatron and LHC, which can shows up as a clear resonant peak in the $t\bar{t}$ invariant mass distribution~\cite{Agashe:2006hk,Lillie:2007yh,Frederix:2007gi,Baur:2008uv}. While current experimental limits depend on the detailed choices of couplings~\cite{Aaltonen:2009tx,Chatrchyan:2012ku,Aad:2012wm}, they nevertheless indicate that COVB with mass below $1$ TeV is severely constrained. \\
\indent It is well known that QCD effects play an important role in $t\bar{t}$ production. The NLO QCD corrections to SM $t\bar{t}$ production, which significantly enhance the $t\bar{t}$ total cross sections, have been calculated for a long time~\cite{Nason:1987xz,Beenakker:1988bq,Beenakker:1990maa}. In the SM, $\afb$ is related to higher order QCD radiation effects, which first appear at $\mathcal{O}(\alpha_s^3)$~\cite{Kuhn:1998kw}. Complete NLO corrections to this observable are not available currently, but calculation based on soft gluon resummation indicates that higher order QCD effects are small~\cite{Ahrens:2011uf}. Recently electroweak corrections to $\afb$ have also been calculated and are found to slightly increase the asymmetry~\cite{Hollik:2011ps}. In the case of massive COVB, its QCD gauge interaction is uniquely determined by its color content, resembling a SM gluon. Therefore, it's reasonable to expect that higher order QCD effects will also have significant impacts on processes mediated by massive COVB, at least at an energy scale comparable to the mass of COVB. This has motivated the model dependent calculation of COVB production by gluon fusion~\cite{Allanach:2009vz} and the model independent~(using dimension-six operators) calculation of $t\bar{t}$ production mediated by COVB~\cite{Bauer:2010iq}. However, a complete NLO analysis of the QCD effects to models with massive COVB in the resonant $t\bar{t}$ region is still absent.~\footnote{While this work was completed, Ref.~\cite{Chivukula:2011ng} appeared, which calculates the singly production of color-octet vector boson in narrow width approximation.}
In this letter, we present the model independent complete NLO QCD corrections to top quark pair production mediated by a general massive COVB,  and show the detailed numerical analysis of top quark pair production, including invariant mass distribution and $\afb$ at the NLO level. We also show that the NLO corrections significantly stabilize the renormalization and factorization scale dependence, as compared to the LO results.

Below, we briefly outline our approach to systematically calculating the NLO QCD effects to processes of COVB production. We consider a model independent massive COVB originated from a broken $SU(3)$ gauge group. The effective Lagrangian for color-octet vector $G_\mu$ in unitary gauge can be written as
\begin{equation}
\label{eq:ga}
  \mathcal{L}_G=-\frac{1}{2}\mathrm{Tr}\,G_{\mu\nu}G^{\mu\nu}
+  M^2_G \mathrm{Tr}\,G_\mu G^{\mu},
\end{equation}
where $a=1,..,8$ are the broken SU(3) ``color'' indices, and $\mu=0,..,3$ are the Lorentz indices. $G_{\mu\nu}\equiv G^a_{\mu\nu}T^a=(\partial_\mu G^a_\nu - \partial_\nu G^a_\mu)T^a$ is the field strength tensor, where $T^a$ is the conventional Gellmann matrix with the normalization $\mathrm{Tr}[T^aT^b]=\frac{1}{2}\delta^{ab}$.  The mass of the COVB is denoted by $M_G$. The QCD color interaction between COVB and SM gluon $A_\mu$ can be easily implemented by changing the ordinary derivative into covariant derivative:
\begin{equation}
\label{eq:two}
  \partial_\mu G^a_\nu \to D_\mu G^a_\nu=\partial_\mu G^a_\nu + g_1 f^{abc} A^b_\mu G^c_\nu,
\end{equation}
where $g_1$ is the coupling constant of QCD. The Lagrangian in Eq.~(\ref{eq:ga}), after the replacement in Eq.~(\ref{eq:two}), is already invariant under the conventional ${\rm SU(3)}_c$ transformation. If desired, NLO QCD calculation can be done with the Feynman rule derived from the above Lagrangian, where unitary gauge is chosen for the broken SU(3) gauge symmetry. However, it is well known that loop calculation in unitary gauge is inconvenient because of the violent ultra-violet (UV) behavior of the propagator. Instead, we choose to carry out the calculation in conventional 't Hooft-Feynman gauge for the broken gauge group. To this end, we separate the longitudinal component~(the would-be Goldstone boson) of the COVB from Eq.~(\ref{eq:ga}), by modifying the mass term in Eq.~(\ref{eq:ga}) as follows,
\begin{equation}
\label{eq:massterm}
 G_\mu(x) \to \tilde{G}_\mu(x) =U(x)^\dagger \left( \frac{i}{g_2}\partial_\mu+G_\mu(x)\right) U(x),
\end{equation}
where we have introduced the $\pi$ field~(the would-be Goldstone field) via $U(x) = e^{i\pi^a(x) T^a/f}$, and the symmetry breaking scale $f=M_G/g_2$, with $g_2$ being the coupling constant
of the broken SU(3) gauge symmetry.
It is easy to check that the mass term is now invariant under the gauge transformation:
\begin{gather}
  G_\mu(x) \to V(x)\left( G_\mu(x) + \frac{i}{g_2} \partial_\mu \right) V(x)^\dagger,\\
U(x) \to V(x) U(x),
\end{gather}
where $V(x)$ is a finite gauge transformation $V(x)=\exp(i\alpha^a_2(x)T^a)$. Similar to unitary gauge, the interaction between $G_\mu$, $\pi$ and QCD gluon is obtained by changing the ordinary derivative into covariant derivative, as in Eq.~(\ref{eq:two}). Using Eqs.~(\ref{eq:two}) and (\ref{eq:massterm}), the classical Lagrangian for COVB turns into
\begin{equation}
\label{eq:ga1}
  \tilde{\mathcal{L}}_G =\left. \left(-\frac{1}{2}\mathrm{Tr}\,G_{\mu\nu}G^{\mu\nu}
+  M^2_G \mathrm{Tr}\,\tilde{G}_\mu \tilde{G}^{\mu}\right)\right|_{\partial_\mu\to D_\mu}.
\end{equation}
Note that the classical Lagrangian $\tilde{\mathcal{L}}_G$ is non-renormalizable, {\it i.e.}, when expanding $U(x)$, it contains operators of dimension larger than $4$, whose coefficients are suppressed by at least one power of $f$. Since we are only interested in low energy QCD effects of COVB rather than its UV-completion, these higher dimensional operators can be safely neglected.
Expanding the mass term in Eq.~(\ref{eq:ga1}), we find that there is a kinetic mixing between the COVB and the would-be Goldstone boson: $ M_G\pi^a(x) \partial_\mu G^a_\mu(x)$. This mixing can be canceled by introducing a gauge fixing term:
\begin{equation}
  \label{eq:fix}
  F_2 = - \frac{1}{2}( \partial^\mu G^a_\mu(x) + M_G \pi^a(x))^2,
\end{equation}
which is similar to the gauge fixing term for the QCD Lagrangian in 't Hooft-Feynman gauge:
\begin{equation}
  \label{eq:fix2}
  F_1 = - \frac{1}{2}( \partial^\mu A^a_\mu(x))^2.
\end{equation}
 The corresponding ghost Lagrangian is given by
\begin{equation}
  \label{eq:ghost}
  \mathcal{L}_g = \bar{u}^i(x) \frac{\delta F_{i}}{\delta \theta_j(x)} u^j(x),
\end{equation}
where $i=1$ for the QCD ghost, and $i=2$ for the broken $SU(3)$ ghost. $\theta_i(x)$ are the infinitesimal gauge transformation parameters for the corresponding gauge group.
The complete Lagrangian for the gauge sector now can be written as
\begin{equation}
\label{eq:full}
  \mathcal{L} = \tilde{\mathcal{L}}_G + \mathcal{L}_{QCD} + F_1 + F_2 + \mathcal{L}_g,
\end{equation}
where $\mathcal{L}_{QCD}$ is the classical QCD Lagrangian.
All the Feynman rules determined by gauge symmetry can then be derived from Eq.~(\ref{eq:full}), and couplings between COVB and SM fermions can be introduced through vector or axial-vector current interaction.  We have checked that this set of Feynman rules are in agreement with those derived in Ref.~\cite{Zhu:2011gd} for the case of RS model~\cite{Randall:1999ee}.

Before continuing, we should define the precise meaning of the NLO QCD corrections in this paper. At LO, there are three parts of contributions to the $t\bar{t}$ cross section in models with massive COVB: the squared SM amplitudes, the interference between the new physics (NP) amplitudes and the SM amplitudes, and the squared NP amplitudes. The NLO QCD corrections in this paper refer to the $\mathcal{O}(\alpha_s)$ corrections to these three parts, respectively. All the relevant Feynman rules can be derived from the effective Lagrangian given in Eq.~(\ref{eq:full}). The results obtained in this way reflect the model independent corrections from QCD interaction. As a final comment, we note that our NLO QCD corrections are the non-Abelian analogy of the QED corrections to $W^\pm$ boson production at hadron collision~\cite{Wackeroth:1996hz}, and the neglected contributions in our calculation are similar to the genuine weak corrections there~\cite{Wackeroth:1996hz}, which can be mostly absorbed into redefinition of
boson and fermion couplings at LO.

Extending the approach shown in our paper~\cite{Zhu:2011gd}, we calculate the one-loop renormalized helicity amplitudes, with the unstable particle treated in the complex mass scheme~\cite{Denner:2006ic}. The loop integrals with complex arguments appearing in the one-loop amplitudes are evaluated with ONELOOP~\cite{vanHameren:2010cp}. The real emission matrix elements are generated by a modified version of MADGRAPH~\cite{Alwall:2011uj}. Soft and~(or) collinear singularities of the real corrections are dealt with by the dipole method~\cite{Catani:2002hc} implemented in the MADDIPOLE package~\cite{Frederix:2010cj}. Throughout our calculation, the pole mass of top quark is chosen as $m_t=173.1$ GeV, and $\alpha_s(M_Z) = 0.118$. LO cross sections are obtained with the CTEQ6L parton distribution  function~(PDF)~\cite{Nadolsky:2008zw} with one-loop running of $\alpha_s$, while NLO cross sections are obtained with the CTEQ6M PDF with two-loop running of $\alpha_s$. In the numerical calculation below we present the results for a benchmark axial-gluon model~\cite{Bai:2011ed} resulting from a simutaneous fit of $t\bar{t}$ total cross section, $\afb$, and dijet resonance searches. In this model, The coupling between the massive COVB and quarks are chosen as
\begin{gather}
  v_q(m_t) = 0, \quad a_q(m_t) = 2.2, \nonumber\\
  v_t(m_t) = 0, \quad  a_t(m_t) = -3.2,
\label{eq:coupling1}
\end{gather}
where $g_1v_{q(t)}$ and $g_1a_{q(t)}$ are the vector and axial coupling of the light quark~(top quark) to the COVB, defined at the scale $m_t$. The evolution for $v_{q,t}(\mu)$ under the change of scale is given by $g_1(\mu)v(\mu)_{q,t}=g_1(\mu_0)v(\mu_0)_{q,t}\left(\alpha_s(\mu)/\alpha_s(\mu_0)\right)^{15/(2\beta_0)}$~\cite{Zhu:2011gd}, where $\beta_0=23/3$ is the first QCD beta-function coefficient for $N_c=3$ and $n_f=5$, and the evolution equation for $a_{q,t}(\mu)$ has the same form. Unless specified, the mass of the COVB is chosen as $2$ TeV. {We note that while such choice of parameters is at the margin of currently allowed parameter space~\cite{Haisch:2011up,Domenech:2012ai,Atre:2012gj}, the qualitative behavior of the  NLO QCD effects are quite general and can be applied in more realistic models. For the calculation of accurate numerical results for other parameter choices, the corresponding results can be directly obtained from our Fortran code.}

We define the NP cross section, $\sigma_{\rm NP}$, as the difference between $t\bar{t}$ cross section in a model with a massive COVB and the SM:
\begin{equation}
\label{eq:sigmanp}
  \sigma_{\rm NP} = \sigma_{\rm SM+NP} - \sigma_{\rm SM},
\end{equation}
where the SM cross sections, including both the $q\bar{q}$- and $gg$-channels, are calculated with the program {\sc MCFM}~\cite{mcfm}.
\begin{figure}[h]
  \begin{center}
  \includegraphics[width=0.5\textwidth]{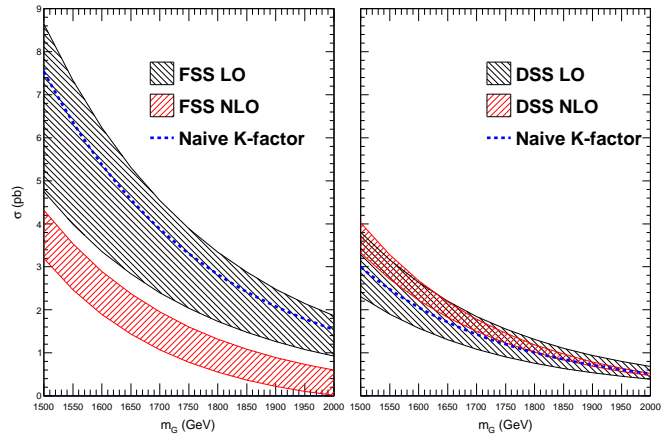}
  \end{center}
  \caption{$\sigma_{\rm NP}$, defined in Eq.~(\ref{eq:sigmanp}), as functions of $M_G$ for FSS~(left) and DSS~(right). The black~(red) bands are the LO~(NLO) uncertainties, estimated by varying the scales around their default values by a factor of two within each scheme. The blue dashed lines are the naive estimates of the NLO effects by simple rescaling of the LO results with the SM K-factor. The couplings are chosen as in Eq.~(\ref{eq:coupling1}).}
  \label{fig:signp}
\end{figure}

In Fig.~\ref{fig:signp}, we plot the NP contribution to the cross section at the LHC with $\sqrt{s}=7$ TeV, as a function of $M_G$. The bands reflect the scale uncertainties estimated by varying the renormalization ($\mu_r$) and factorization ($\mu_f$) scales around their default values by a factor of $2$ with $\mu_r=\mu_f$. We present the results in two benchmark schemes, namely the fixed scale scheme~(FSS), i.e. the scales are fixed to be $m_t$, and the dynamical scale scheme~(DSS), i.e. the scales are set to be the invariant mass of top quark pair $m_{t\bar{t}}$. We find that the NLO QCD effects are 
large and small in the FSS and the DSS, respectively, for our choice of parameters and the naive estimate of the NLO effects by simple rescaling of the LO results with the SM NLO K-factor is not appropriate. It is also clear that the inclusion of NLO QCD effects strongly reduces the theoretical uncertainty of the NLO cross section in the both scheme, compared with the LO ones.
\begin{figure}[h!]
  \begin{center}
  \includegraphics[width=0.5\textwidth]{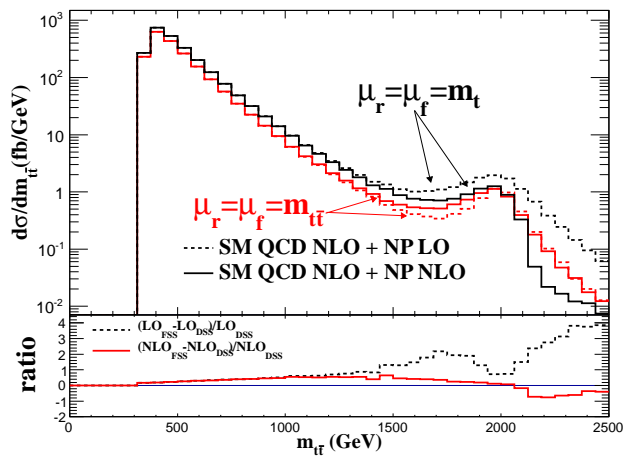}
  \end{center}
  \caption{LO~(dashed lines) and NLO~(solid lines) invariant mass distribution of the top quark pair at the LHC with $\sqrt{s}=7$ TeV are shown in the upper panel. In the bottom panel the differences between two different schemes at the LO and NLO are shown. In both panels, red curves and black curves correspond to FSS and DSS, respectively. The couplings are chosen as in Eq.~(\ref{eq:coupling1}).}
  \label{fig:dis1}
\end{figure}

Fig.~\ref{fig:dis1} gives the LO and NLO invariant mass distributions of the top quark pair at the LHC with $\sqrt{s}=7$ TeV, including contribution from SM LO and NLO top quark pair production. It can be seen that the NLO results in the DSS has relatively small corrections compared with the LO one,
while the NLO QCD corrections in the FSS significantly change the shape of the LO curve, leading to the reduction of the width of the resonance. This is due to the fact that the NLO width can be expressed analytically as
\begin{equation}
\label{eq:width}
  \frac{\Gamma_{\rm NLO}(\mu)}{\Gamma_{\rm LO}(\mu)}=\left[1+\frac{\alpha_s(\mu)}{\pi}
\left( \frac{167}{12}-\pi^2 - \frac{15}{4}\ln\frac{M^2_G}{\mu^2}\right)\right].
\end{equation}
From Eq.~(\ref{eq:width}),
it is obvious that the width of COVB is significantly reduced at NLO in FSS, while the large logarithmic contributions are absent in DSS. The predictions in the two schemes at NLO are close to each other, while at LO they show large difference, as shown in the lower panel of Fig.~\ref{fig:dis1}. The difference between the two schemes reflects the uncertainties of the theoretical prediction. Hence, the NLO result leads to a smaller theoretical uncertainties in $m_{t \bar t}$ distribution, which could improve the accuracy of extracting the theory parameters of NP models from comparing to experimental data. We also note that similar conclusion holds in other cases, e.g., KK gluon in RS model.

From the calculations of the total cross section and the invariant mass distribution, we see that the perturbation expansion may break down when FSS is chosen because the COVB resonance mass is usually about an order of magnitude larger than the top quark mass, leading to very large logarithms in the perturbation series. In contrast, the scale in DSS is related to the dynamics of the process and thus is a more reasonable choice.

\begin{figure}[h]
  \begin{center}
  \includegraphics[width=0.5\textwidth]{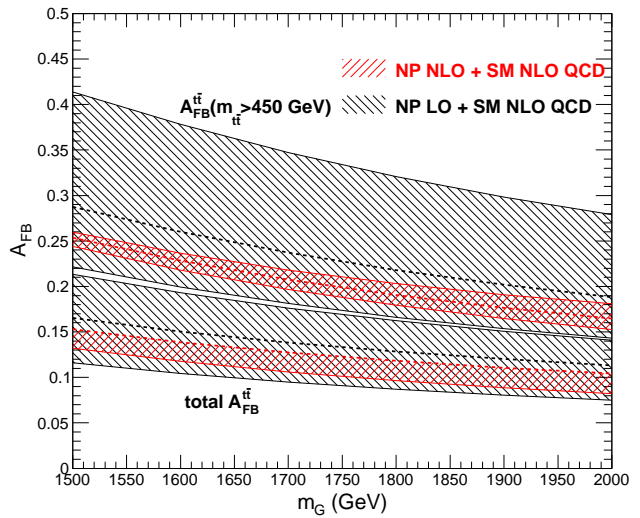}
  \end{center}
  \caption{NP contributions to the $A_{\rm FB}$ in the $t\bar{t}$ center of mass frame as a function of the mass of the COVB at the Tevatron with $\sqrt{s}=1.96$ TeV. In the upper lines only the $A_{\rm FB}$ in the large invariant mass region, $m_{t\bar{t}}>450$ GeV, are plotted, while in the bottom total $A_{\rm FB}$ are shown. The bands reflect scale dependence, estimated by varying the scales around their central values by a factor of $2$. The predicitions for $\mu=m_t$ are depicted in dashed lines. The couplings are chosen as in Eq.~(\ref{eq:coupling1}).}
  \label{fig:afb}
\end{figure}

Fig.~\ref{fig:afb} shows the LO and NLO contributions to the $A_{\rm FB}$ as a function of the mass of the COVB at the Tevatron with $\sqrt{s}=1.96$ TeV in the center of mass frame of $t\bar{t}$ pair. The results are given for both the total asymmetry~(bottom bands) and the asymmetry in the large invariant mass region~(upper bands), $m_{t\bar{t}}>450$ GeV, respectively. The bands in Fig.~\ref{fig:afb} reflect the scale uncertainties of the theoretical predictions, which are obtained by simutaneously varying $\mu_r$ and $\mu_f$ in the numerator and denominator of $\afb$ around $m_t$ by a factor of $2$ with $\mu_r=\mu_f=\mu$. Here we choose the FSS, because the SM results $A_{FB}^{tot}=0.158\pm0.075$ is also obtained at the scale $\mu=m_t$. In both cases, the most significant effects of the NLO corrections are the reduction of scale dependences. It is worth pointing out that the numerator of $\afb$ is dominated by NP contributions, while the denominator is dominated by SM $t\bar{t}$ cross section. Therefore at LO the large scale dependence in the numerator, which is mainly from scale dependence of NP coupling $v_{q,t}(\mu)$ and $a_{q,t}(\mu)$, can not be canceled by the corresponding scale dependence in the denominator. The predictions for $\afb$ at $\mu=m_t$ are depicted in Fig.~\ref{fig:afb} as dashed lines. It can be seen that the NP contributions at NLO only reduce $A_{\rm FB}$ by $3-4\%$.

\begin{figure}[h]
  \begin{center}
  \includegraphics[width=0.5\textwidth]{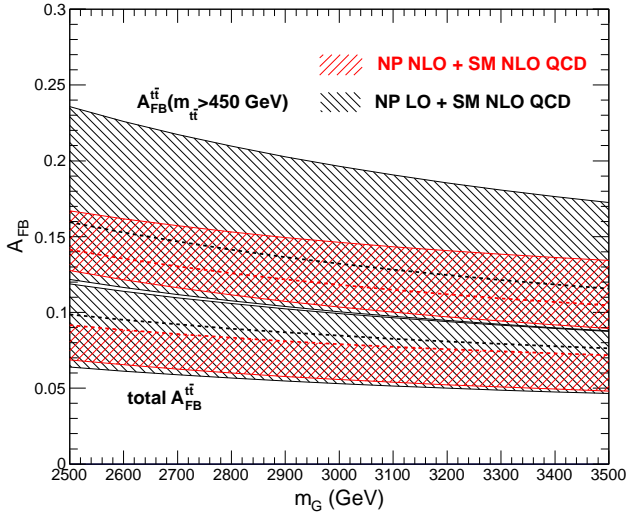}
  \end{center}
  \caption{NP contributions to the $A_{\rm FB}$ in the $t\bar{t}$ center of mass frame as a function of the mass of the COVB at the Tevatron with $\sqrt{s}=1.96$ TeV. In the upper lines only the $A_{\rm FB}$ in the large invariant mass region, $m_{t\bar{t}}>450$ GeV, are plotted, while in the bottom total $A_{\rm FB}$ are shown. The bands reflect scale dependence, estimated by varying the scales around their central values by a factor of $2$. The predicitions for $\mu=m_t$ are depicted in dashed lines. The couplings are chosen as in Eq.~(\ref{eq:coupling2}).}
  \label{fig:afb2}
\end{figure}

{Figs.~\ref{fig:signp}, \ref{fig:dis1}, and \ref{fig:afb} show the cross sections as functions of $M_G$, invariant mass distribution and the $A_{\rm FB}$ as a function of $M_G$ for a COVB with axial couplings, respectively. For the $A_{\rm FB}$, it would also be interesting to illurstrate the results for a COVB with both vectorial and axial couplings. For this purpose, we consider another set of couplings for a COVB with $M_G=3$ TeV as follows,
\begin{gather}
  v_q(m_t) = -1, \quad a_q(m_t) = -3, \nonumber\\
  v_t(m_t) = -3, \quad  a_t(m_t) = -1.
\label{eq:coupling2}
\end{gather}
The results for this set of couplings are shown in Fig.~\ref{fig:afb2}. Again we can see that the NP contributions at NLO only reduce $A_{\rm FB}$ by a small amount for default scale choice~($\mu=m_t$). }

\begin{figure}[h!]
    \begin{center}
    \includegraphics[width=0.45\textwidth]{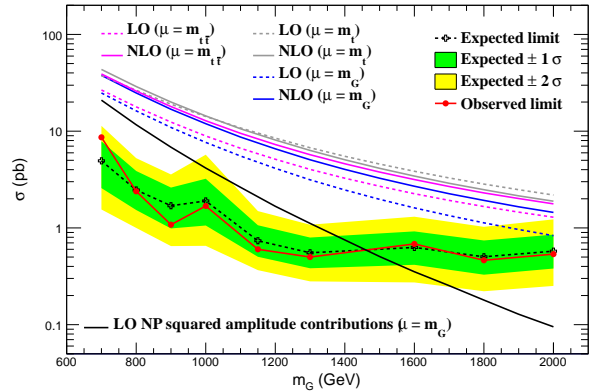}
    \end{center}
    \caption{LO and NLO predictions for $\sigma_{\rm NP}$ in a specific RS model~\cite{atlasconf}. The expected and observed limits on cross section are extracted from the Ref.~\cite{atlasconf}. }
    \label{fig:totcs}
  \end{figure}

As a further application of our results, we plot in Fig.~\ref{fig:totcs} the NP cross section, $\sigma_{\rm NP}$, for a specific RS model considered in search for resonance decaying into top quark pairs by ATLAS collaboration~\cite{atlasconf}, where the left-handed and right-handed couplings~(modulo $g_s$) between KK gluon and light quarks are chosen as
\begin{equation}
  c_{L,q}(M_G)=c_{R,q}(M_G)=-0.2,
\end{equation}
and similarily the couplings between KK gluon and top quark are chosen as
\begin{equation}
  c_{L,t}(M_G)=1,\quad c_{R,t}(M_G)=4.
\end{equation}
 Also plotted in Fig.~\ref{fig:totcs} are the experimental exclusion limit extracted from the same paper. Our exact theoretical predictions are given for three different scale choices, $\mu=m_t$, $\mu=M_G$ and $\mu=m_{t\bar{t}}$, at both LO and NLO. It can be seen that the NLO QCD effects are moderate in the $\mu=m_{t\bar{t}}$ case, while they are large in both the $\mu=m_t$ and $\mu=M_G$ cases. We also plot in Fig.~\ref{fig:totcs} the LO results~(the black solid line) with only the contributions from NP squared amplitudes, {\it i.e.}, without the interference with the SM amplitudes. It is clear that considering the NP squared amplitudes contributions alone obviously underestimates the NP cross sections, and including the interference contributions is necessary for reliable predictions.
We note that while the mass limit for KK gluon is very different at LO for the three kinds of scale choices, their differences are significantly reduced at  NLO. Hence, the NLO results can be used for precise extraction of mass limit for KK gluon.

Finally, we give the numerical results of the NP cross section for the same set of parameters as in Fig.~\ref{fig:totcs}, with $M_G$ fixed to be $1.5$ TeV. The results are presented for three different scale choices, i.e. $\mu=m_t$, $\mu=M_G$ and $\mu=m_{t\bar{t}}$ with $\mu_r=\mu_f=\mu$. The LO~(NLO) results are calculated with CTEQ6L~(CTEQ6M) in Table~\ref{tab:cteq} and with MSTW2008LO~(MSTW2008NLO) in Table~\ref{tab:mstw}, respectively. In the LO calculations, we only estimate the theoretical uncertainties by varying the renormalization and factorization scales around their central values by a factor of $2$. In the NLO calculations, we also consider the PDF uncertainties, and the first error is scale uncertainty and the second is PDF uncertainty. It can be seen from Tables~\ref{tab:cteq} and \ref{tab:mstw} that scale uncertainties are reduced from LO to NLO in all three scale choices, and the differences between different scale choices are also reduced when going from LO to NLO. Moreover, the predictions of total cross sections for CTEQ and MSTW PDFs are obviously different at LO, but consistent within the PDF uncertainties at NLO.

\begin{table}[h!]
  \begin{center}
      \begin{tabular}{c|ccc}
        \hline \hline
        & $\mu=m_t$ &$\mu=M_G$& $\mu = m_{t\bar{t}}$
        \\ \hline
        $\sigma^{\rm LO}_{\rm NP}$ [pb] & $4.57^{+1.43}_{-1.14}$ &
        $1.98^{+0.57}_{-0.42}$ & $2.69^{+0.85}_{-0.61}$
        \\ \hline
       $\sigma^{\rm NLO}_{\rm NP}$ [pb]& $4.13^{+0.38+0.22}_{-0.38-0.22}$ & $3.26^{+0.41+0.17}_{-0.45-0.17}$  &$3.77^{+0.31+0.20}_{-0.39-0.20}$
        \\ \hline \hline
      \end{tabular}
  \end{center}
  \vspace{-3ex}
  \caption{\label{tab:cteq}The NP cross sections for top quark pair production mediated by KK gluon at the LHC~($\sqrt{s}=7$~TeV) at the LO~(CTEQ6L) and NLO~(CTEQ6M). Theoretical uncertainties are estimated by varying the renormalization and factorization scales around their central values by a factor of $2$.}
\end{table}

\begin{table}[h!]
  \begin{center}
      \begin{tabular}{c|ccc}
        \hline \hline
        & $\mu=m_t$ & $\mu=M_G$ & $\mu = m_{t\bar{t}}$
        \\ \hline
        $\sigma^{\rm LO}_{\rm NP}$ [pb] & $6.20^{+2.44}_{-1.69}$ &
        $2.44^{+0.79}_{-0.56}$ & $3.48^{+1.24}_{-0.87}$
        \\ \hline
       $\sigma^{\rm NLO}_{\rm NP}$ [pb]& $4.88^{+0.43+0.19}_{-0.43-0.19}$ & $3.51^{+0.45+0.15}_{-0.44-0.15}$ & $4.08^{+0.14+0.17}_{-0.44-0.17}$
        \\ \hline \hline
      \end{tabular}
  \end{center}
  \vspace{-3ex}
  \caption{\label{tab:mstw}The NP cross sections for top quark pair production mediated by KK gluon at the LHC~($\sqrt{s}=7$~TeV) at the LO~(MSTW2008LO) and NLO~(MSTW2008NLO). Theoretical uncertainties are estimated by varying the renormalization and factorization scales around their central values by a factor of $2$.}
\end{table}

In conclusion, we have presented a theoretical framework for systematically calculating NLO QCD effects to various experimental observables in models with massive COVB in a model independent way. Specifically, we show the numerical results for the NLO QCD corrections to total cross sections, invariant mass distribution and $\afb$ of top quark pairs produced by mediating a massive COVB. Our results show that, for our choice of parameters, the NLO corrections of the NP cross section are much larger in the FSS than in the DSS, but the NLO QCD calculations in the dynamical scale scheme is more reasonable than the fixed scheme, and the naive estimate of the NLO effects by simple rescaling of the LO results with the SM NLO K-factor is not appropriate. Moreover, for invariant mass distribution, we find that the NLO QCD corrections change the width of the resonant particle in the DSS slightly, and the differences in the results using FSS and DSS are reduced after including NLO QCD effects.

We would like to thank Liang Dai and Jun Gao for collaboration on early stage of this project, and Qing-Hong Cao, Michele Petteni, Bernd Stelzer, and Jing Shu for helpful discussion.
This work was supported in part by the National Natural Science
Foundation of China, under Grants No.11021092 and No.10975004. C.P.Y
acknowledges the support of the U.S. National Science Foundation
under Grand No. PHY-0855561.
\bibliography{color_octet}{}
\end{document}